\documentclass[twocolumn,superscriptaddress,showpacs,pre,aps]{revtex4}
\usepackage{graphicx}
\usepackage{subfigure}

\begin{document}

\title{Coiling Instabilities of Multilamellar Tubes}

\author{C. D. Santangelo}%
\email{santa@mrl.ucsb.edu}
\affiliation{Department of Physics, University of California, Santa Barbara, CA 93106}
\author{P. Pincus}%
\affiliation{Department of Physics, University of California, Santa Barbara, CA 93106}
\affiliation{Department of Materials, University of California, Santa Barbara, CA 93106}
\affiliation{Program in Biomolecular Science and Engineering, University of California, Santa Barbara, CA 93106}
\date{\today}%

\begin{abstract}
Myelin figures are densely packed stacks of coaxial cylindrical bilayers that are unstable to the formation of coils or double helices.  These myelin figures appear to have no intrinsic chirality.  We show that such cylindrical membrane stacks can develop an instability when they acquire a spontaneous curvature or when the equilibrium distance between membranes is decreased.  This instability breaks the chiral symmetry of the stack and may result in coiling.  A unilamellar cylindrical vesicle, on the other hand, will develop an axisymmetric instability, possibly related to the pearling instability.
\end{abstract}


\maketitle

\section{Introduction}
\begin{figure}
	\resizebox{3in}{!}{\includegraphics{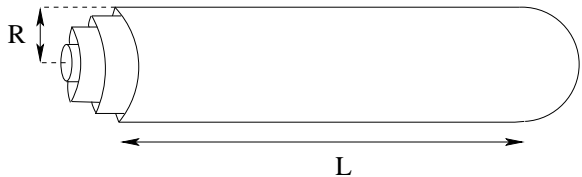}}
	\caption{\label{fig1} A myelin figure is a multilamellar cylindrical stack of bilayers.  Typical radii are on the order of 10 $\mu m$.}
\end{figure}

Amphiphilic molecules in water self-assemble into a variety of structures, including micelles, bilayer vesicles, and stacks of bilayer membranes.  When a dehydrated lump of amphiphile is brought into contact with water, it will develop a large number of tubular structures which grow to lengths many times larger than their radii~\cite{myelingrowth}.  These tubes, called {\it myelin figures}, are densely-packed, nested stacks of coaxial cylindrical bilayers~\cite{myelinsection}.  Typically the outer radius of such a stack is on the order of tens of microns and the inner radius can be as small as or smaller than $0.2 \mu m$~\cite{weizmann1} (see FIG.~\ref{fig1}).  Myelin figures tend to be straight over lengths many times longer than their radii, but eventually bend or fold back on themselves~\cite{myelinhelix}.  Though they appear to have no intrinsic chirality, they are observed to form coils or double helices~\cite{myelinhelix,weizmann1,weizmann2}.

Coiling has also been observed in cylindrical stacks of binary mixtures~\cite{myelincoiling}, and egg-yolk phosphatidylcholine~\cite{myelinhelix}.  In the first system, $Ca^{2+}$ ions were added to myelin figures formed from a binary mixture of cardiolipin and phosphatidylcholine.  Above a critical concentration of ions, tightly-packed, single helices were formed.  Coiling was also observed when the system was hydrated with a solution of $Ca^{2+}$ ions.  In the latter system, the myelin figures would be straight initially but form helices and other structures after they had been growing for several minutes.  It was presumed that the observed coiling in both systems was due to surface adhesion.  However, coiling has also been observed recently in another system of myelin figures hydrated with a polymer solution in which it was demonstrated that surface adhesion played a negligiable role in the coiling mechanism~\cite{weizmann1,weizmann2}.  Coiling in this system was attributed to an interaction between the polymer and the membranes which resulted in a local spontaneous curvature.  In~\cite{weizmann2}, a model was developed in which the polymer is allowed to diffuse freely along the bilayer surfaces.  Once the polymer concentration is above a critical concentration, the straight myelin figure becomes unstable to one which is maximally coiled.

Unilamellar tubes in the presence of the same polymer, on the other hand, are known to exhibit a {\it pearling instability}~\cite{weizmannpearling}.  Here, the tubes are formed first and a drop of polymer solution is added later.  In the early stages of the instability, the tubes develop a periodic oscillation along their length which leads, eventually, to the formation of a string of spheres.  A different pearling instability has also been observed in unilamellar tubes excited with optical tweezers~\cite{opticaltweezerpearling} and has been attributed to induced surface tension~\cite{dynamicalpearlingtheory1,dynamicalpearlingtheory2}, and unilamellar tubes with oil incorporated into the bilayer~\cite{oilpearlinginstability}.

In this paper, we calculate the bending and interaction energies for undulations of a cylindrical multilamellar stack. We then study the instabilities generated by perturbing the stack in various ways: (1) we add spontaneous curvature to all the bilayers, and (2) we decrease the interaction equilibrium distance between the layers after the stacks have been formed.  We show that the most unstable mode (neglecting dynamics) breaks the chiral symmetry of the cylinders and introduces a twisting undulation, which may be related to coiling.  In contrast, spontaneous curvature on a unilamellar cylindrical vesicle results in periodic oscillation along the cylinder axis, as in the pearling instability.

A similar instability in a smectic-A liquid crystal confined within a cylindrical channel with homeotropic boundary conditions was studied in~\cite{cylindricalsmectic}.  The boundary conditions cause the smectic layers to arrange themselves into co-axial cylinders much like myelin figures.  The presence of the confining walls changes the nature of the instability, however, since the outer layer of a myelin figure is not fixed to be a cylinder.  Also, those authors consider only undulations which vary along the length of the cylinder which explicitely prohibits instabilities that break chiral symmetry.

The remainder of the paper is organized as follows.  In section II we calculate the total energy of a cylindrical multilamellar stack to second order in arbitrary undulations of the radius.  In section III we demonstrate the existence of the instabilities.  In section IV, we discuss the nature of the instabilities.  Finally, we summarize in section V. 

\section{Energy of a cylindrical stack}
\subsection{Geometry}
We will assume the bilayers in the cylindrical stack to be equally spaced\footnote{The assumption of equally spaced cylindrical layers is not made in reference~\cite{cylindricalsmectic} (a smectic-A with cylindrical layers), or reference~\cite{deswelling} (myelin figures that have been deswelled nonuniformly).}, though the actual distribution of layers is unknown.  At length scales larger than molecular lengths, it is appropriate to model a bilayer membrane as a two dimensional surface.  There are two equivalent representations of the membrane surfaces that we will use.

On one hand, we can introduce local coordinates on the cylinders $(\theta,z)$ and a vector function $\vec{X}(\theta,z)$ which gives the position in space of any point on the membrane.  The $n^{th}$ membrane in the stack can then be described by $\vec{X}(\theta,z)=r_n(\theta,z) (\cos(\theta) \hat{x} + \sin(\theta) \hat{y}) + z \hat{z}$ where $r_n(\theta,z) = r_0$ is independent of $\theta$ and $z$ for a cylinder of radius $r_0$.

We generalize this to $r_n(\theta,z)=a n + h_n(\theta,z)$ where $a$ is the spacing between the layers, and $h_n(\theta,z)$ is an undulation of the cylinder radius.  Since the membranes in the stack are very close together, the undulations of the layers are severely constrained and $h_n(\theta,z)$ should be slowly varying radially.  Therefore, to simplify our calculations, we specialize to the case where $h_n(\theta,z)$ is independent of $n$.

The induced metric, $g_{\alpha \beta}$ ($\alpha,\beta = \theta,z$), is defined by $g_{\alpha \beta} = \partial_{\alpha} \vec{X} \cdot \partial_{\beta} \vec{X}$ where $\partial_{\theta} = \frac{\partial}{\partial \theta}$ and $\partial_{z} = \frac{\partial}{\partial z}$.  The differential surface area element is $dA = d\theta dz \sqrt{g}$ where $g \equiv \det g_{\alpha \beta}$.  The mean curvature and Gaussian curvature are defined to be the trace and determinant respectively of the curvature tensor, $K_{\alpha \beta}$, given by, $K_{\alpha \beta} \hat{n} = D_{\alpha} \partial_{\beta} \vec{X}$ where $D_{\alpha}$ is the covariant derivative.  Then $H = \arrowvert \triangle \vec{X} \arrowvert$ where $\triangle=g^{\alpha \beta} D_\alpha \partial_{\beta}$ is the Laplace-Beltrami operator acting on a scalar.  In two dimensions, $\int dA \det K_{\alpha \beta}$ is a topological invariant.

An alternative way to represent the layers is with a function $\phi(r,\theta,z)=r-u(r,\theta,z)$.  The $n^{th}$ membrane is then given by the solution to the equation $\phi(r,\theta,z)=a_0 n$ for some $a_0$.  It is straightforward to show that $u(r,\theta,z) = \frac{(a - a_0) r +a_0 h(\theta,z)}{a}$ describes a cylinder with a radius given by $r_n(\theta,z) = a n + h(\theta,z)$.  In this representation, we write the unit normal as $\hat{n} = \frac{\nabla \phi}{\arrowvert \nabla \phi \arrowvert}$.  This gives the unit normal as a function of $(r,\theta,z)$ so we use $\hat{n}(r_n(\theta,z),\theta,z)$ for the unit normal on the $n^{th}$ membrane.  Finally, we write $\tilde{H}(r,\theta,z) = \nabla . \hat{n}$ and $H_n(\theta,z) = \tilde{H}(r_n(\theta,z),\theta,z)$ gives the mean curvature on the $n^{th}$ membrane.  This representation is similar to that used for smectic liquid crystals (see, for example, ~\cite{deGennes}).

\subsection{Bending Energy}
The theory of the bending energy of amphiphilic membranes is well-established for small curvatures~\cite{bendingenergy}.  The bending energy is given by,
\begin{equation}
E_B = \frac{\kappa}{2} \int dA (H - H_0)^2 + \bar{\kappa} \int dA K
\end{equation}
where $H$ is the mean curvature, $K = \det K_{\alpha \beta}$ is the Gaussian curvature, and $H_0$ is the spontaneous curvature.  Spontaneous curvature can arise from an asymmetry between the two monolayers of the membrane. Here we will assume that $H_0 = 0$.  Since $\int dA K$ is a topological invariant and we do not allow topology change, we suppress this term.

We assume that the surface area per molecule is constant.  Though myelin figures are nonequilibrium structures, we will also assume that their rate of growth, $0-0.3 \mu m/s$~\cite{weizmann1}, is slow enough to allow us to treat them as being in equilibrium at any given time.  These assumptions together give us the constraint that the total surface area of the layers should remain constant.  To impliment this constraint, we add an effective surface tension of the form $\sigma \int dA$ where $\int dA$ is the surface area of the membrane and the surface tension, $\sigma$, is actually a Lagrange multiplier chosen to enforce the constraint.

This single constraint, however, is not enough to ensure that a cylindrical membrane is an extremum of the bending energy.  If we neglect the endcap energy (which is reasonable for long cylinders), the mean curvature for a cylinder of radius $\rho$ is $1/\rho$.  Thus, the bending energy of a cylinder can always be lowered by increasing the radius, and we can always do so while conserving area by decreasing the length.  An additional contribution to the free energy is necessary for a cylindrical membrane to be stable.

Since the mechanism which stabilizes a myelin figure is still poorly understood, there are a number of possible approaches we can take to this problem.  On the one hand, Sakurai and Kawamura~\cite{myelingrowth} suggest that a cylindrical stack can be stabilized by the interaction energy, which forbids the layers from increasing their separation.  This still allows for the possibility of increasing all the radii without changing their separations.  A second possibility is that of osmotic pressure which keeps the volume between the layers constant.  In~\cite{Helfrichstability}, the authors use osmotic pressure to stabilize a cylindrical vesicle and their results have a simple generalization to cylindrical stacks.  However, for mathematical simplicity, we choose to solve this by adding a linear tension (of the form $- t L$).  By choosing the linear tension and surface tension properly, we can ensure that a cylindrical stack of any length and distribution of layers is an extremum of the energy.  In section V, we comment on how conserving the volume between the layers changes our results.

Finally, the effective energy of each membrane layer, without interactions between the layers  and with $H_0 = 0$, is given by,
\begin{equation}
e_B = \sigma \int d^2x \sqrt{g} + \frac{\kappa}{2} \int d^2x \sqrt{g} H^2 - t L
\end{equation}

It will be convenient to use the Fourier series representation for the function $h(\theta,z)$.  Since the cylinder length is much longer than the radius, we will neglect the contribution of the ends of the stack to the energy and will impose periodic boundary conditions along the cylinder axis.  This allows us to write $h(\theta,z)=\sum_{m j} b_{m j} e^{i m \theta + i q_j z}$ where $q_j=\frac{2 \pi j}{L}$, $L$ is the cylinder length, and both $m$ and $j$ are integers.  The reality of $h(\theta,z)$ imposes the condition $b_{m j}^* = b_{-m -j}$ on the Fourier coefficients where $^*$ signifies the complex conjugate.  The coefficient $b_{0 0}$ represents changes of radius of all the bilayers.  This coefficient is redundant with the layer separation $a$ and the core radius.

We calculate $\sigma$ and $t$ for each layer by imposing the conditions that the first derivative of the bending energy of a straight cylinder with respect to both radius and length should vanish.  This gives us $\sigma = \frac{\kappa}{2 \rho^2}$ and $t = \frac{2 \pi \kappa}{\rho}$, where $\rho$ is the radius of the cylinder.  Expanding the bending energy to second order in $h(\theta,z)$ and its derivatives and using these values for $\sigma$ and $t$ results in the expression,
\begin{eqnarray}\label{eq:layerbend}
e_B & = & \pi L \kappa \sum_{m j} \arrowvert b_{m j} \arrowvert^2 \bigg\{ \frac{(m^2-1)^2}{\rho^3} + {}\nonumber\\
& & {} + \rho q_j^4 + \frac{2 m^2 q_j^2}{\rho} \bigg\}
\end{eqnarray}
for the bending energy of each bilayer, where $\rho=a n$ is the $n^{th}$ membrane radius.

Notice that the first order variation of the bending energy with respect to the undulations $h(\theta,z)$ vanishes since there are no terms in the bending energy linear in $b_{m j}$.  Therefore, we can conclude that a straight cylinder is an extremum of this bending energy.  Also, notice that the undulation mode with $j=0$ and $m=1$ has zero bending energy.  Since the shape change associated with this mode causes a change in the mean curvature of higher than second order, it is typically associated with translation of the cylinder normal to the long axis.

Finally, to find the total bending energy we sum over the membranes.  Since they are very dense, we go over to the continuum limit and replace $\sum_n$ with $\frac{1}{a} \int_{r_c}^R d\rho$ where $R$ is the radius of the stack, and $r_c$ is the radius of the core.  The final result for the bending energy is,
\begin{eqnarray} \label{eq:stackbend}
E_B & = & \frac{\pi L \kappa}{a} \sum_{m j} \arrowvert b_{m j} \arrowvert^2 \bigg\{ \frac{(m^2-1)^2}{2 r_c^2} + {} \nonumber\\
& & {} + \frac{R^2}{2} q_j^4 + 2 m^2 q_j^2 \ln(R/r_c) \bigg\}
\end{eqnarray}
where we have used the fact that $r_c \ll R$.

\subsection{Interaction Energy}
The largest contribution to the interaction energy comes from nearest-neighbor membranes.  Consider the shortest curve between the two membranes which is normal to both surfaces.  If the membrane separation and size of undulations are small, this curve is nearly straight and, for any point $(\theta,z)$ on one membrane with normal curve passing through it, the interaction energy gets its dominant contribution from the point on the other membrane which lies along the curve.  Let $\tilde{a}$ be the length of the normal curve between the two membranes, $a$ be the separation of the layers with no undulations, and $a_0$ be the equilibrium separation of the layers.  For simplicity, we will assume that $a$ and $a_0$ are constant (in particular, they are independent of $n$).

To find $\tilde{a}(\theta,z)$, we represent the membrane surfaces as solutions to the equation $a_0 n = \phi(r,\theta,z) = r - u(r,\theta,z)$.  Then, noting that $\oint_C \frac{1}{\tilde{a}} \hat{n} = 0$ in the absence of defects and that $\hat{n}=\frac{\nabla \phi}{\arrowvert \nabla \phi \arrowvert}$, we can write $\tilde{a}(\theta,z) = \frac{a}{\arrowvert \nabla \phi \arrowvert} \approx a [1 - (\frac{\partial u}{\partial r} - \frac{1}{2} (\nabla u)^2)]$

We can then expand the interaction energy in powers of $(\tilde{a} - a_0)/a_0$ to second order.  The first order term vanishes since $a_0$ is the equilibrium separation, so the interaction energy density is  $\frac{B}{2}[\frac{\partial u}{\partial r} - \frac{1}{2} (\nabla u)^2]^2$ where $B$ is the bulk modulus.  This must be integrated over the surface of each membrane in the layer.  This interaction energy is similar to the familiar de Gennes elastic term for the free energy of a smectic liquid crystal~\cite{deGennes}.  However, our bulk modulus $B$ has units of energy per area rather than energy per volume as it does in a liquid crystal.

To properly choose $\sigma$ and $t$, we should take into account the interaction energy of the membranes.  Doing so gives a correction to our value for $\sigma$ at nonzero $\gamma = \frac{a - a_0}{a_0}$.  In fact, this gives $\sigma = \frac{\kappa}{2 \rho^2} - \frac{B}{2} \frac{\gamma^2 (1+\gamma/2)^2}{(1+\gamma)^4}$ where $\rho = (1+\gamma) a_0 n$ is the radius of the $n^{th}$ membrane.

Using our value for $u(r,\theta,z)$, the resulting interaction energy per layer is,
\begin{eqnarray}
e_I & = & \frac{\pi L B}{(1+\gamma)^4} \Big\{  - \gamma (1+\gamma/2) \sum_{m j} \arrowvert b_{m j} \arrowvert^2 \big[ \frac{m^2}{\rho} + {} \nonumber\\
& & {} + q_j^2 \rho \big] \Big\}
\end{eqnarray}

If there are no undulations, the interaction energy is zero even for nonzero $\gamma$.  This comes about because of the correction to $\sigma$.  Without this correction, a cylindrical stack with $a \ne a_0$ will not be an extremum of the energy.

We finally note that if $\gamma=0$, there is no contribution from the interaction energy to second order.  This is because we are not allowing the undulation of the radius, $h(\theta,z)$, to vary from bilayer to bilayer so, to second order, they remain equidistant.

\section{Instabilities}
\subsection{Spontaneous Curvature}
Adding the terms,
\begin{equation} \label{eq:spontcurv}
E_c = - \kappa \int d^2x \sqrt{g} H H_0 + \frac{\kappa}{2} \int d^2x \sqrt{g} H_0^2
\end{equation}
 to the bending energy of each layer gives each membrane a spontaneous curvature, $H_0$.  In order to conserve surface area the surface tension, $\sigma$, will adjust itself to cancel the second term exactly.  Without changing the linear tension, however, there will appear terms in the total energy that are linear in the $b_{m j}$ arising from the first term of equation (\ref{eq:spontcurv}).  This corresponds to an instability that {\it decreases} the radii of the lamellae.  However, the interaction energy of the stack requires that the change in radius of the layers be independent of the radii of the layers.  Conservation of surface area implies that the change in length, $\delta L$, be related to the change in radius, $\delta R$, by the expression $\delta L = - \delta R L / \rho$.

Since $\delta R < 0$, we see that the length of each layer is increased.  However, the change of length is smaller for the layers with larger radii.  The endcap of the inner layers will, therefore, be able to reach the endcaps of the outerlayers very early in this instability.  This will prevent any large changes of length in the cylindrical stack.  Fixing the length of the cylindrical stack can be accomplished by adjusting the linear tension to be $t=\frac{2 \pi \kappa}{\rho} - 2 \pi \kappa H_0$.  With these changes, terms that are first order in the $b_{m j}$ vanish.  Expanding the resulting energy to second order in the undulations results in the following additional term to equation (\ref{eq:layerbend}),
\begin{equation}
e_c = -2 \pi L \sum_{m j} \arrowvert b{m j} \arrowvert^2 \bigg\{ \frac{C m^2}{\rho^2} + D q_j^2 \bigg\}
\end{equation}
with $C=0$ and $D=\kappa H_0$.  Approximating the sum over membranes with an integral, the total stack energy is calculated to be $\frac{a}{2 \pi L} E = \sum_{m j} \arrowvert b_{m j} \arrowvert^2 E_{m j}$, where,
\begin{eqnarray}
E_{m j} & = & \frac{-C m^2}{r_c} - D q_j^2 R + \frac{\kappa q_j^4 R^2}{4} + {} \nonumber\\
& & {} + \frac{\kappa (m^2-1)^2}{4 r_c^2} + \kappa m^2 q_j^2 \ln(R/r_c)
\end{eqnarray}

Since $L$ is large, $q_j$ is well-approximated as a continuous variable, $q$.  For $m=0$, we have,
\begin{equation}
E_{0 q} = - \kappa H_0 q^2 R + \frac{\kappa q^4 R^2}{4} + \frac{\kappa}{4 r_c^2}
\end{equation}

Minimizing this to find the most unstable mode gives $q^2 = \frac{2 H_0}{R}$ with an onset threshold of $H > \frac{1}{2 r_c}$.  For $m=1$, we have,
\begin{equation}
E_{1 q} = -\kappa H_0 q^2 R + \frac{\kappa q^4 R^2}{4} + \kappa q^2 \ln (R/r_c)
\end{equation}

This expression is negative whenever $q^2 < q_c^2 = \frac{4 H_0}{R} - \frac{4}{R^2} \ln(R/r_c)$ and $H_0 > \frac{1}{R} \ln(R/r_c)$.  This threshold is much smaller than the threshold for the $m=0$ instability.  Finally, larger values of $m$ can be analyzed like the $m=0$ case, yielding even larger thresholds for instability.  Notice that $q_c^2$ is proportional to $H_0 - H_{0 c}$, so the critical wavelength is very long for spontaneous curvatures just above $H_{0 c}$.  The most unstable wavelength can be found by minimizing $E_{1 q}$.  Doing so gives the most unstable mode to be $q^2 = 2 \frac{H_0}{R} - {2}{R^2} \ln(R/r_c)$.

We can do a similar analysis on a unilamellar membrane by using the energy for a single layer.  The procedure above establishes the existence of an instability when $H_0 > 1/\rho$ where $\rho$ is the cylinder radius for both $m=0$ and $m=1$.  The $m=1$ instability starts with long wavelength modes, however, while the $m=0$ instability has an onset wavelength of $2 \pi \rho$ and is expected to dominate.

This result qualitatively agrees with Ou-Yang and Helfrich's calculation of instabilities in cylindrical vesicles~\cite{Helfrichstability} stabilized by osmotic pressure.  One crucial difference, however, is that the $m=1$ mode never goes unstable in their calculation.  The discrepancy must be in our choice of linear tension over osmotic pressure to stabilize the cylindrical vesicle.

\subsection{Decreasing the interaction equilibrium distance}
We now consider the situation when the equilibrium distance between the membranes is decreased.  This requires special care because the mode which relaxes the equilibrium distance by reducing the radii uniformly and increasing the length has been neglected by assuming $h(\theta,z)$ is independent of radius.  This relaxation mode carries a linear dependence on the radius, $\rho$, and will need to be considered explicitely.

There are two effects which compete against the complete relaxation of the interaction energy through this mode.  The first is the curvature energy which resists decreasing the radii of any of the membrane layers.  The second is the endcap energy which resists changes of length.  The stack will be unable to completely relax its interaction energy because of these effects, and the resulting separation between the layers, $a$, will be larger than the new equilibrium distance, $a_0$.  This is mathematically equivalent to a swelled stack, which for a flat, lamellar phase results in a Helfrich-Hurault type instability~\cite{helfrichhurault}.

For a cylindrical stack, this results in an additional contribution to equation (\ref{eq:layerbend}) of,
\begin{equation}
e_I = -2 \pi L \sum_{m j} \arrowvert b_{m j} \arrowvert^2 \bigg\{\frac{C m^2}{\rho} + D q_j^2 \rho \bigg\}
\end{equation}
where $C = D = \frac{B}{2} \frac{\gamma (1 + \gamma/2)}{(1+\gamma)^4}$.  Summing over all membranes in the stack gives a total energy $E = \frac{2 \pi L}{a_0} \sum_{m j} \arrowvert b_{m j} \arrowvert^2 E_{m q_j}$, where,
\begin{eqnarray}
E_{m q_j} & = & D m^2 \ln(R/r_c) + \frac{D}{2} q_j^2 R^2 + \frac{\kappa (m^2-1)^2}{4 r_c^2}\\ \nonumber
& & {} + \frac{\kappa R^2}{4} q_j^4 + \kappa m^2 q_j^2 \ln(R/r_c)
\end{eqnarray}

We again study the stability of the energy for different choices of $m$ while taking $q_j$ to be continuous.  For $m=0$, we have,
\begin{equation}
E_{0 q} = -\frac{D q^2 R^2}{2} + \frac{\kappa}{4 r_c^2} + \frac{\kappa R^2 q^4}{4}
\end{equation}

This equation gives a most unstable mode of $q^2=\frac{D}{\kappa}$ and an instability whenever $D>\frac{\kappa}{R r_c}$.  For $m=1$, we have,
\begin{equation}
E_{1 q} = - D \ln (R/r_c) - \frac{D q^2 R^2}{2} + \frac{\kappa R^2 q^4}{4} + \kappa q^2 \ln (R/r_c)
\end{equation}
which is negative when $q^2 < q_c^2 = \frac{D}{\kappa} - \frac{2}{R^2} \ln(R/r_c) + \sqrt{\frac{D^2}{\kappa^2} + \frac{4 [\ln(R/r_c)]^2}{R^4}}$.  Thus, the reality of $q$ requires that $D > D_c = 0$, or $\gamma > 0$.  Again, this threshold is smaller than the $m=0$ threshold and the thresholds for instability for $m>1$ are even higher than the $m=0$ case.  We see then that, as $D$ is increased, the first mode to destabilize is the one with $m=1$ and $q^2 \approx \frac{D}{\kappa}$ for small $\gamma$.  We find the most unstable wavenumber to be $q=\frac{D}{\kappa} - \frac{2}{R^2} \ln(R/r_c)$.

The critical wavelength, $2 \pi / q_c = (2 \pi \lambda \sqrt{2})/\sqrt{\gamma}$ where $\lambda = \sqrt{B/\kappa}$ is the smectic penetration length, can be quite short.  A 0.1 percent reduction in equilibrium distance results in a critical wavelength of approximately $280 \lambda$.  If $\lambda$ is on the order of the layer spacing and there are several hundred layers in the cylindrical stack, the critical wavelength will be on the order of a tube radius.  A one percent reduction in equilibrium distance results in a critical wavelength smaller than a single tube radius.

\section{Coiling of myelin figures}

\begin{figure}
	\subfigure[] {
		\label{fig2:subfig:a}
		\resizebox{!}{3in}{\includegraphics{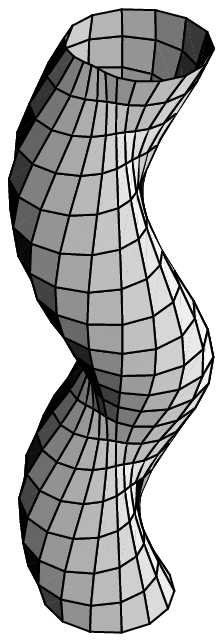}}}
	\subfigure[] {
		\label{fig2:subfig:b}
		\resizebox{!}{3in}{\includegraphics{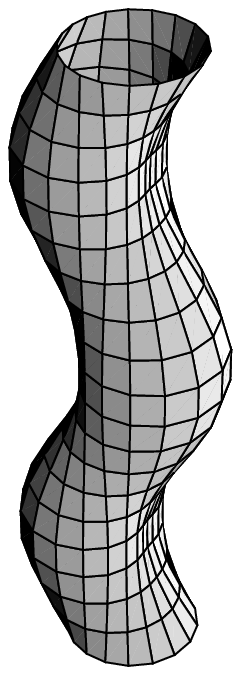}}}
\caption{\label{fig2}(a) A three dimensional representation of a twisting undulation. (b) A three dimensional representation of a writhing undulation.}
\end{figure}

Whenever $q>0$, there are two types of behavior that can result, which we refer to as twisting (or coiling) and writhing.  Twisting can be represented by a model undulation of the form $h(\theta,z) = \alpha \cos(\theta + q z)$, for example, which breaks the chiral symmetry of the cylinders.  Writhing, on the other hand, can be represented by $h(\theta,z)=\alpha \cos(\theta) \cos(q z)$ (see FIG.~\ref{fig2}).  Fourier coefficients which give twisting for $m=1$ and some choice of $q$ are $b_{1 q} = \alpha/2$ and $b_{-1 q}=0$ with the remaining coefficients zero or set by the reality condition on $h(\theta,z)$.  In the case of writhing, we have $b_{1 q} = \alpha /4$ and $b_{-1 q}=\alpha/4$.  The normalization factors of $2$ and $4$ are necessary so that the amplitudes of both undulations are equal to $\alpha$.  Now we can write the energy for both instabilities in the form,
\begin{eqnarray}
E_{twisting} = E_{m q} \frac{\alpha^2}{2} \\
E_{writhing} = E_{m q} \frac{\alpha^2}{4}
\end{eqnarray}

So we see that whenever $E_{m q} < 0$, the most unstable case is twisting and not writhing.  Since the $m=1$ mode moves the center axis of the cylinders in the stack, we speculate that this mode may result in coiling as seen in experiments.

In both types of instabilities discussed, the first mode to go unstable is the mode with $m=1$ and $q=0$.  We expect that, when dynamics is taken into account, this $q=0$ mode will grow more slowly than all other wavelengths.  Thus, the dominant wavelength will be at some $q > 0$.  It is not clear, however, how dynamics will affect whether the stack twists or writhes, and whether the behavior will lead to coiling.

\section{Summary and Discussion}
In this paper, we considered a nested stack of cylindrical, interacting membranes.  We studied the instabilities generated by perturbing the stack in two ways: (1) we add spontaneous curvature to all the bilayers, and (2) we decrease the interaction equilibrium distance between the layers after the stacks have been formed.  In both cases, we find that the first modes to go unstable are long wavelength modes which break the rotational symmetry of the cylinder.  We find that the most unstable modes break chiral symmetry and may be related to the coiling which is observed in myelin figures.

We should stress that we have studied only two in a larger class of instabilities.  If we imagine perturbing the system by adding a term to the free energy that prefers local undulations of the layers, we are constrained, by dimensional analysis, to write down,
\begin{equation}
E_P = 2 \pi L \sum_{m j} \arrowvert b_{m j} \arrowvert^2 (- C m^2 \rho^\nu - D q_j^2 \rho^{\nu+2})
\end{equation}
to quadratic order in $b_{m j}$, $m$, and $q_j$.  The units of $C$ and $D$ (assumed the same) determine the value of $\nu$.  As we have shown, a chiral symmetry breaking instability will occur when $\nu=-1$ or $\nu=-2$.  Different mechanisms might then give rise to similar instabilities, depending only on the dimensions of their coefficients and the general condition that they result in local undulations.  We expect, therefore, that there may be many instabilities of nested cylindrical stacks that lead to similar behavior.

The instability resulting from decreasing the layer equilibrium distance bears some resemblence to an instability due to deswelling a myelin figure discovered by Chen, {\textit et al.}~\cite{deswelling,mackintosh}.  They consider myelin figures whose outer layers only have been deswelled.  In this case, the in-plane elasticity of the layers becomes relevant.  They find that deswelling a myelin figure in this way results in an axisymmetric instability ($m = 0$).

It is important to note that allowing $\gamma$ or $H_0$ to vary with radius can change the results completely.  We see this already if we consider adding spontaneous curvature only to the outermost layers of the stack.  In that case, the additional energy will be approximately proportional to the energy of a cylindrical vesicle.  As we have already seen, an energy of this form has a short wavelength axisymmetric instability ($m=0$) and a long wavelength helical instability ($m=1$), both at the same threshold.

We have already mentioned that our results for a spontaneous curvature-induced instability in a cylindrical vesicle disagree with those of~\cite{Helfrichstability}, which uses osmotic pressure rather than linear tension to stabilize the vesicles.  This sensitivity to our choice of constraints carries over to a cylindrical stack, also.  A cylindrical stack with fixed volume between the layers which acquires spontaneous curvature on all of its layers will not develop an instability at $m=1$ as in the case of a cylindrical vesicle.  The first mode to destabilize must then be $m=0$.  We have checked, however, that using osmotic pressure on a cylindrical stack whose equilibrium distance has been decreased still results in an instability with $m=1$ and onset $\gamma>0$, but does change the value of $q_c$.

\begin{acknowledgements}
We would like to thank F.C. MacKintosh and T. Lubensky for stimulating and useful discussions.  This work was supported by the National Science Foundation under Award No. DMR-9972246, and by the MRL Program of the National Science Foundation under Award No. DMR00-80034.
\end{acknowledgements}

\end{document}